\documentclass[conference]{IEEEtran}
\IEEEoverridecommandlockouts

\usepackage{cite}
\usepackage{hyperref}
\usepackage{amsmath,amssymb,amsfonts}
\usepackage{algorithmic}
\usepackage{graphicx}
\usepackage{textcomp}
\usepackage{xcolor}

\def\BibTeX{{\rm B\kern-.05em{\sc i\kern-.025em b}\kern-.08em
    T\kern-.1667em\lower.7ex\hbox{E}\kern-.125emX}}
\begin{document}

\title{Surface Material Dependence in Powder Triboelectric Charging\\
\thanks{Funding was provided by the Engineering and Physical Sci-
ences Research Council through the Centre for Doctoral Training in Aerosol
Science (no. EP/S023593/1) and materials provided by the Process Hazards
Department at Syngenta Huddersfield.}
}

\author{
\IEEEauthorblockN{1\textsuperscript{st} Tom F. O'Hara}
\IEEEauthorblockA{\textit{Faculty of Science and Engineering} \\
\textit{University of Bristol}\\
Bristol, England \\
tom.ohara@bristol.ac.uk}
\and
\IEEEauthorblockN{2\textsuperscript{nd} Ellen Player}
\IEEEauthorblockA{\textit{Process Hazards} \\
\textit{Syngenta}\\
Huddersfield, England \\
email address or ORCID}
\and
\IEEEauthorblockN{2\textsuperscript{nd} Graham Ackroyd}
\IEEEauthorblockA{\textit{Process Hazards} \\
\textit{Syngenta}\\
Huddersfield, England \\
email address or ORCID}
\and
\IEEEauthorblockN{2\textsuperscript{nd} Peter J. Caine}
\IEEEauthorblockA{\textit{Process Hazards} \\
\textit{Syngenta}\\
Huddersfield, England \\
email address or ORCID}
\and
\IEEEauthorblockN{5\textsuperscript{th} Karen L. Aplin}
\IEEEauthorblockA{\textit{Faculty of Science and Engineering} \\
\textit{University of Bristol}\\
Bristol, England \\
karen.aplin@bristol.ac.uk}
}

\maketitle

\begin{abstract}
Triboelectric charging of granular materials against container walls is a critical yet poorly understood phenomenon affecting many industrial powder handling processes. Charge accumulation can cause material flow disruptions, adhesion issues, and pose serious safety risks, such as providing ignition sources for dust cloud explosions. This study investigates particle-wall charging behaviour for materials with varying size, shape, and electrical resistivity. Aluminium surfaces are used as a reference case, followed by analysis of labradorite, an analogue for volcanic ash, to examine charging interactions with various wall materials. Finally, the triboelectric response of industrially relevant materials used in flexible intermediate bulk containers, including both lined and unlined variants, is assessed. Results show that stainless steel surfaces generate the least charge, while the presence of insulating polyethene liners increases charging significantly. These findings contribute to a deeper understanding of triboelectric charging mechanisms in powder transfer operations and inform safer industrial practice.
\end{abstract}


\section{Introduction}

Many industrial processes rely heavily on granular materials, which can become significantly charged through triboelectric (frictional) mechanisms that remain poorly understood \cite{matsusaka_triboelectric_2010, lacks_long-standing_2019}. Charge accumulation can disrupt material flow, lead to unwanted adhesion, reduce process efficiency, and, in severe cases, trigger dust cloud explosions via electrostatic discharge \cite{pingali_use_2009, glor_electrostatic_2005}. Such charging is frequently observed in operations such as pneumatic conveying, sieving, and fluidised bed processing \cite{wilms_ml_2024, deng_electrostatic_2023, murtomaa_electrostatic_2003}.

Particles can charge with other surfaces, such as container walls, referred to as particle-wall charging in this work, or with other particles. The extent to which particles become charged and retain charge is influenced by both material properties and environmental conditions. Key material properties include electrical conductivity, surface roughness, wettability, and particle size \cite{heinert_decay_2022, jantac_triboelectric_2025, grunebeck_size_2024}, while external factors such as temperature and relative humidity are also known to affect charging behaviour significantly \cite{xu_experimental_2023, cruise_triboelectric_2023}.

Despite considerable research, particle–wall triboelectric charging remains difficult to predict, especially in complex, real-world environments such as the pneumatic conveying of powders \cite{alfano_computational_2021}. This poses challenges for risk mitigation, particularly in powder transfer operations where electrostatic discharges can act as ignition sources \cite{puttick_avoidance_2008}. As a result, many facilities adopt nitrogen inertisation to displace oxygen and reduce the risk of explosion, despite the high energy demand and environmental cost associated with gas generation \cite{choi_experimental_2015, aneke_potential_2015}.

One area of particular concern is powder transfer into or from reactor vessels, especially when using flexible intermediate bulk containers (FIBCs). These containers vary widely in material composition and are employed in a range of transfer systems \cite{glor_electrostatic_2005, ebadat_testing_1996}. Gravity-fed drops are often preferred due to their simplicity and low environmental impact, but the triboelectric charging that occurs during these operations is not well characterised. Industry guidance typically limits drop heights to three metres to reduce charging risk, though this practice is based more on precaution than on robust experimental evidence \cite{puttick_avoidance_2008}.

This study investigates particle–wall triboelectric charging for a range of materials differing in particle size, shape, and electrical resistivity, with aluminium surfaces used as a reference case in Section \ref{Aluminium Reference Drops}. A major component of the volcanic ash samples, identified as labradorite through X-ray diffraction (XRD), is used as a simplified analogue to explore charging interactions with various wall materials in Section \ref{Delivery Tube Material Investigation}. Finally, Section \ref{FIBC Lining} evaluates the triboelectric response of industrial wall materials found in flexible intermediate bulk containers, comparing unlined and lined configurations.

\begin{figure*}[t!]
\centerline{\includegraphics[width=1.0\textwidth]{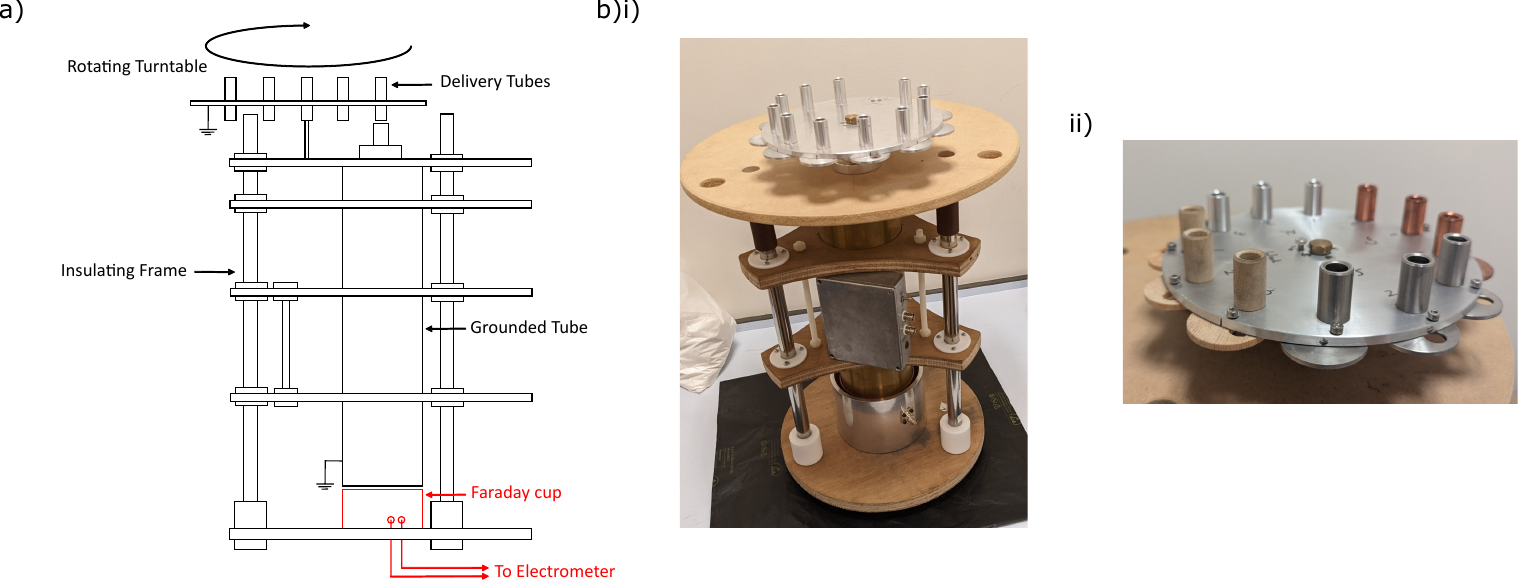}}
\caption{a) The schematic for the experimental setup of the charge drop apparatus used for powder drops, adapted from previous works \cite{ohara_faraday_2025, houghton_triboelectric_2013}. b) Images of the experimental setup for the i) powder drop rig with aluminium delivery tubes and ii) the custom turntable designed with aluminium, stainless steel, copper and wooden delivery tubes close up.}
\label{Experimental Rig}
\end{figure*}

\section{Methods}
\label{Methods}

To investigate particle charging with aluminium and other materials, a custom experimental setup was developed, as shown in Figure \ref{Experimental Rig}. For each material, the release mechanism comprised a delivery tube and a sliding trap door constructed from the material under study. Larger-scale measurements, involving Flexible Intermediate Bulk Containers (FIBCs) with and without polyethylene liners (1.75 kg of powder), were conducted by dropping powder from the FIBC into a stainless steel hopper.

The primary measurement technique employed Faraday cups to measure changes in voltage ($\Delta V$), which were converted into changes in charge ($\Delta Q$) using the known capacitance ($C$) of the system, according to the relation $\Delta V = \Delta Q / C$. Capacitance values were determined to be 130 pF and 630 pF for the small and larger-scale setups, respectively. These values were obtained by applying a continuously varying voltage ($dV/dt$) and calculating capacitance from the measured current ($I$) using $I = C \cdot dV/dt$ \cite{aplin_self-calibrating_2001, houghton_triboelectric_2013, ohara_faraday_2025}.

For voltage measurement, the Faraday cups were connected via triaxial BNC connectors to a \href{https://ilg.physics.ucsb.edu/Courses/RemoteLabs/docs/Keithley6514manual.pdf}{Keithley 6514 electrometer}. To minimise noise during small-scale measurements, short and rigid connections were used. For the larger FIBC and hopper measurements, flexible cables were acceptable due to the improved signal-to-noise ratio \cite{keithley_low_1998}. The electrometer interfaced with a \href{https://www.ni.com/docs/en-US/bundle/usb-6211-specs/page/specs.html}{USB-6210 Data Acquisition (DAQ) device} from National Instruments, controlled via \href{https://www.ni.com/en/support/downloads/software-products/download.labview.html}{LabVIEW$^{\circledR}$ 2024-Q1}, to record data. Environmental conditions, including relative humidity and temperature, were monitored using a \href{https://datasheet.octopart.com/386-Adafruit-Industries-datasheet-81453130.pdf}{DHT11 sensor} connected to an \href{https://docs.arduino.cc/resources/datasheets/A000067-datasheet.pdf}{Arduino Mega 2560}. Although temperature and humidity were not controlled, their variations were recorded during experiments and are outlined in section \ref{Aluminium Reference Drops}. Post-processing of the data was implemented in \href{https://docs.python.org/3/}{Python 3.12.2}.

Scanning Electron Microscopy (SEM) images were obtained using Hitachi TM3030Plus and VEGA3 TESCAN tabletop microscopes. Particle size distributions were measured with a Malvern Mastersizer 3000 using both aero- and liquid-dispersion methods. Powder resistivity measurements were conducted using an Agilent 4339B High Resistance Meter connected to a powder resistivity cell with dimensions 10 cm by 1 cm (area $A$), and a sample length $L$ of 1 cm. Resistivity ($\rho$) was calculated using an applied voltage ($V$) of up to 1000 V and the resulting current ($I$) according to $\rho = (V/I) \cdot (A/L)$.

\section{Results and Discussion}

\subsection{Aluminium Reference Drops}
\label{Aluminium Reference Drops}

\begin{figure*}[t!]
\centerline{\includegraphics[width=1.0\textwidth]{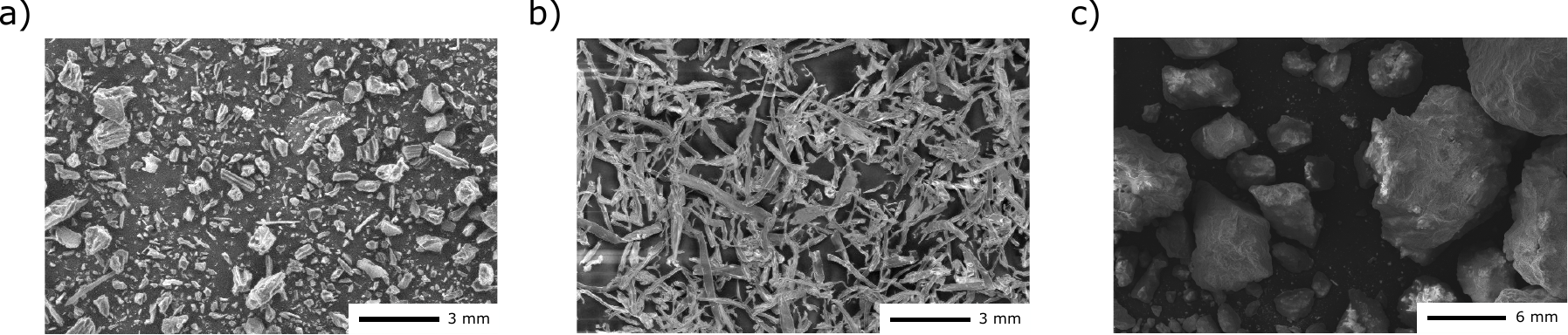}}
\caption{The SEM images of samples of a) carbon black, b) cellulose fibres, and c) volcanic ash from Fuego, Guatemala. a) and b) were taken with the Hitachi TM3030Plus, and c) was taken with the VEGA3 TESCAN tabletop microscope. Each respective scalebar is shown.}
\label{SEM}
\end{figure*}

Initially, three materials: cellulose fibres, volcanic ash from Fuego, and carbon black were dropped using the apparatus and methodology described in Section \ref{Methods}, with aluminium delivery tubes and a dropping mechanism as the reference surface. These materials were selected for their varied surface chemistries, particle shapes, and sizes (shown in Figure \ref{SEM} and Table \ref{resistivity}), making the results with aluminium more broadly applicable.

\begin{figure}[b!]
\centerline{\includegraphics[width=0.5\textwidth]{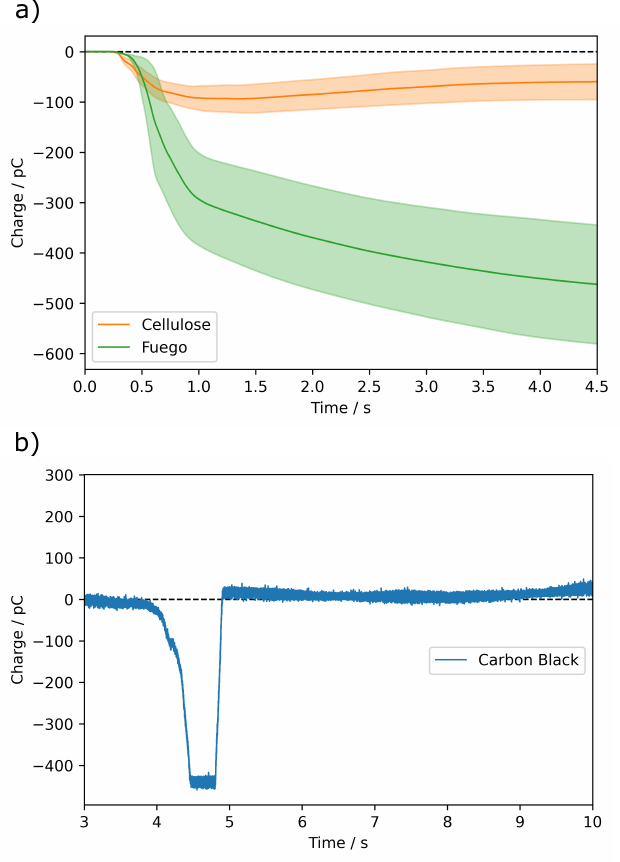}}
\caption{a) Averaged Faraday cup charge traces over time for cellulose fibres and volcanic ash from Fuego, Guatemala, with error regions indicating two standard errors of the mean at each time step. b) Faraday cup trace for a single release of carbon black.}
\label{Charge Traces}
\end{figure}

The Faraday cup traces in Figure \ref{Charge Traces}a show that Fuego ash produced the highest overall charge, with cellulose and ash displaying distinct trace shapes. The cellulose trace initially drops negative before returning toward positive values, suggesting that negatively charged particles land first, followed by positively charged ones. This pattern indicates a significant contribution from particle–particle charging, as discussed in previous work \cite{ohara_faraday_2025}. The slower rise of the cellulose trace also reflects the reduced fall speed of the particles, due to the fibrous structure of cellulose (Figure \ref{SEM}b), creating greater air resistance.

Table \ref{resistivity} shows the resistivity measurements for materials conditioned at 20\% RH and 20~$^\circ$C, as required by industrial fire safety standards, as well as unconditioned values measured under typical lab conditions (37 ± 10\% RH and 21 ± 4~$^\circ$C). Cellulose shows only a slight decrease in resistivity when unconditioned, while carbon black is significantly less insulating, especially without conditioning.

Despite the variation in shape and composition, the Malvern Mastersizer 3000 data show that the materials all have particle sizes of a similar order of magnitude. The crushed labradorite was sieved between 125 and 250~$\mu$m, consistent with its measured mode of 163~$\mu$m.

The distinctive shape of the carbon black trace in Figure \ref{Charge Traces}b likely reflects its low resistivity and partial conductivity. While charge initially appears to build up by induction as the powder falls, it is subsequently dissipated through the Faraday cup, returning the signal to baseline. The other powers are considered insulting and thus retain their charge for a longer period. 

\begin{table}[ht]
\centering
\caption{Resistivity ($\rho$) values of materials conditioned at 20\% RH and 20~$^\circ$C for 24 hours, compared to unconditioned samples at 37 ± 10\% RH and 21 ± 4~$^\circ$C. Modal volumetric particle sizes ($d_p$) were measured using the Malvern Mastersizer.}
\begin{tabular}{|l|c|c|c|}
\hline
Material & $\rho$ / $\Omega\cdot$m & Unconditioned $\rho$ / $\Omega\cdot$m & $d_p$ / $\mu$m \\
\hline
Cellulose    & $9.5 \times 10^{11}$ & $8.9 \times 10^{11}$ & 55\\
Carbon Black & $3.8 \times 10^{10}$ & $9.1 \times 10^{8}$ & 115\\
Fuego Ash    & - & - & 81\\
Labradorite  & - & - & 163\\
\hline
\end{tabular}
\label{resistivity}
\end{table}

\subsection{Delivery Tube Material Investigation}
\label{Delivery Tube Material Investigation}

To investigate the influence of delivery tube and release mechanism material on powder charging, four materials were selected: aluminium (Al), copper (Cu), stainless steel (SS), and wood. The metals are commonly used in industrial processes, while wood, often described as triboelectrically neutral, occupies a central position in the triboelectric series \cite{sun_functionalized_2021}.

\begin{figure}[t!]
\centerline{\includegraphics[width=0.5\textwidth]{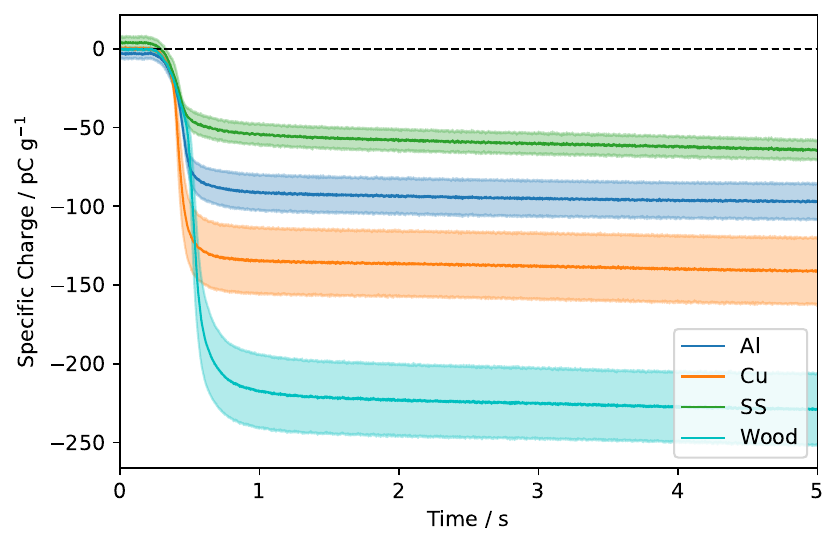}}
\caption{The averaged Faraday cup charge traces over time for the samples of powered labradorite sieved between 125 and 250 $\mu$m. The material of the delivery tube and release machanism is varied as aluminium (Al), copper (Cu), stainless steel (SS), and wood. The error region shown indicates one standard error on the mean at each time step.}
\label{Charging Drops Rig Material Investigation}
\end{figure}

Figure \ref{Charging Drops Rig Material Investigation} shows that net charging, defined here as the difference between the start and end of each averaged Faraday cup trace, is greatest with wood and smallest with stainless steel. The high charging observed with wood is consistent with its insulating nature and high surface roughness, despite its nominal triboelectric neutrality \cite{sun_functionalized_2021, antony_untreated_2025}. In contrast, stainless steel’s low charging is likely due to its high electrical conductivity, work function, and smooth surface.

These results suggest that stainless steel is the most effective of the tested materials for reducing particle–wall charging in systems such as pneumatic conveyors. However, this conclusion is based on tests using labradorite, and the outcomes may differ for powders with different surface chemistries. Further investigation is needed to determine the generality of these findings.

\subsection{FIBC Lining}
\label{FIBC Lining}

The effect of insulating liners on powder charging in flexible intermediate bulk containers (FIBCs) was assessed by comparing conductive bags with and without polyethylene liners. These liners are commonly used for contamination control or moisture protection but may hinder electrostatic dissipation. To evaluate the charge reduction offered by unlined conductive FIBCs, polypropylene granules were used as a simplified, low-risk powder substitute. Powdered labradorite was not used due to the high mass required (1.75~kg), which posed an unacceptable respiritory safety risk.

\begin{figure}[b!]
\centerline{\includegraphics[width=0.45\textwidth]{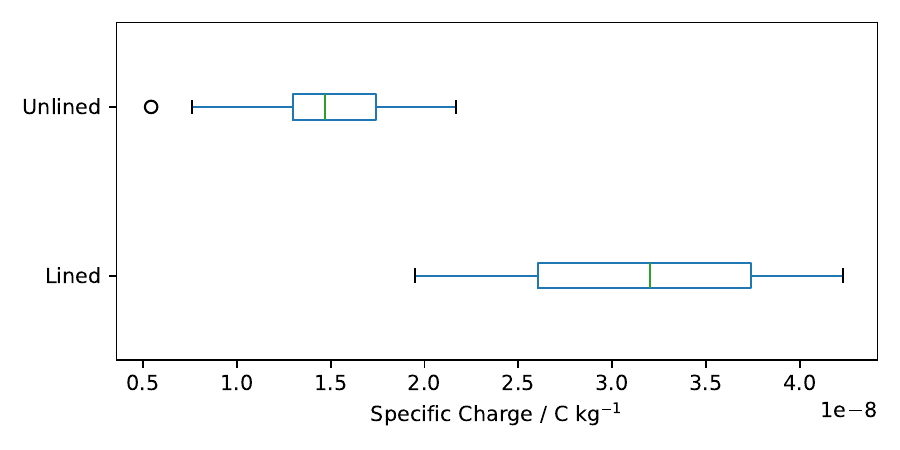}}
\caption{Faraday cup measurements of total charge following powder drops from conductive FIBCs, with and without polyethylene liners. The powder used was polypropylene granules (1.75~kg per drop). Boxes show the interquartile range and median; whiskers span 1.5 times the interquartile range.}
\label{FIBC Boxplot}
\end{figure}

As shown in Figure \ref{FIBC Boxplot}, the presence of the insulating liner roughly doubles the net charge on the powder, indicating that the liner inhibits charge dissipation. Removing the liner in conductive bags significantly reduces charging and supports their use as a mitigation strategy. However, a substantial residual charge remains, so existing ignition risk controls should not be removed. Electrostatic discharge elimination alone is not a sufficient basis of safety for this type of powder handling.

\section{Conclusions}
\label{Conclusions}

This study investigated triboelectric charging of powders against a range of surface materials using controlled Faraday cup experiments at both small and large scales. Powders with varying size, shape, and electrical resistivity were tested, with aluminium used as a reference surface. Volcanic ash and its analogue labradorite were employed to examine particle–wall interactions, while polypropylene granules enabled safe large-scale testing within flexible intermediate bulk containers (FIBCs).

Stainless steel surfaces consistently produced the lowest particle–wall charging, suggesting their suitability for reducing electrostatic risk in powder transfer processes. In contrast, wooden surfaces generated higher charging, likely due to their insulating nature and surface roughness. The addition of polyethylene liners in conductive FIBCs also led to a marked increase in charging, indicating that such insulating layers hinder charge dissipation.

These results highlight the importance of surface material selection in managing electrostatic hazards in powder handling systems. While the trends observed with labradorite and polypropylene provide valuable insight, further work is needed to determine how widely these findings apply across different powders and environmental conditions.

Overall, this work contributes new experimental evidence to support safer, more cost-effective industrial design. Future research should investigate dynamic particle–wall interactions, the influence of humidity at scale, and a broader range of material combinations under representative process conditions.

\section*{CRediT authorship contribution statement}
Tom F. O'Hara: conceptualisation, methodology, formal analysis, validation, data curation, writing – original draft, writing – review and editing, project administration. Ellen Player supervision, writing – review and editing. Graham Ackroyd: supervision, resources. P. J. Caine: investigation. Karen L. Aplin: supervision, writing – review and editing, funding acquisition.

\section*{Data Availability}
All data supporting the findings of this study are publicly available via the Materials Data Facility \cite{ohara_surface_2025}. The code used for post-processing and figure generation is released under the GPL-3.0 license \cite{tom_ohara_vb22224surface-material-dependence--powder-triboelectric-charging_2025}.

\section*{Conflict of Interest}
The authors declare no conflicts of interest.

\section*{Acknowledgments}
The authors thank Gregory Marsden for conducting the XRD measurements and Benedict Brown for providing the SEM image of Fuego volcanic ash. We also acknowledge the broader hazards team at Syngenta for their support with equipment and facilities. This work was supported by the Engineering and Physical Sciences Research Council through the Centre for Doctoral Training in Aerosol Science (grant no. EP/S023593/1), and materials were provided by the Process Hazards Department at Syngenta Huddersfield.

\bibliographystyle{IEEEtran}
\bibliography{references}

\begin{thebibliography}{10}
\providecommand{\url}[1]{#1}
\csname url@samestyle\endcsname
\providecommand{\newblock}{\relax}
\providecommand{\bibinfo}[2]{#2}
\providecommand{\BIBentrySTDinterwordspacing}{\spaceskip=0pt\relax}
\providecommand{\BIBentryALTinterwordstretchfactor}{4}
\providecommand{\BIBentryALTinterwordspacing}{\spaceskip=\fontdimen2\font plus
\BIBentryALTinterwordstretchfactor\fontdimen3\font minus \fontdimen4\font\relax}
\providecommand{\BIBforeignlanguage}[2]{{%
\expandafter\ifx\csname l@#1\endcsname\relax
\typeout{** WARNING: IEEEtran.bst: No hyphenation pattern has been}%
\typeout{** loaded for the language `#1'. Using the pattern for}%
\typeout{** the default language instead.}%
\else
\language=\csname l@#1\endcsname
\fi
#2}}
\providecommand{\BIBdecl}{\relax}
\BIBdecl

\bibitem{matsusaka_triboelectric_2010}
\BIBentryALTinterwordspacing
S.~Matsusaka, H.~Maruyama, T.~Matsuyama, and M.~Ghadiri, ``\BIBforeignlanguage{en}{Triboelectric charging of powders: {A} review},'' \emph{\BIBforeignlanguage{en}{Chemical Engineering Science}}, vol.~65, no.~22, pp. 5781--5807, Nov. 2010. [Online]. Available: \url{https://linkinghub.elsevier.com/retrieve/pii/S0009250910004239}
\BIBentrySTDinterwordspacing

\bibitem{lacks_long-standing_2019}
\BIBentryALTinterwordspacing
D.~J. Lacks and T.~Shinbrot, ``\BIBforeignlanguage{en}{Long-standing and unresolved issues in triboelectric charging},'' \emph{\BIBforeignlanguage{en}{Nature Reviews Chemistry}}, vol.~3, no.~8, pp. 465--476, Jul. 2019. [Online]. Available: \url{https://www.nature.com/articles/s41570-019-0115-1}
\BIBentrySTDinterwordspacing

\bibitem{pingali_use_2009}
\BIBentryALTinterwordspacing
K.~C. Pingali, S.~V. Hammond, F.~J. Muzzio, and T.~Shinbrot, ``\BIBforeignlanguage{en}{Use of a static eliminator to improve powder flow},'' \emph{\BIBforeignlanguage{en}{International Journal of Pharmaceutics}}, vol. 369, no. 1-2, pp. 2--4, Mar. 2009. [Online]. Available: \url{https://linkinghub.elsevier.com/retrieve/pii/S0378517308008636}
\BIBentrySTDinterwordspacing

\bibitem{glor_electrostatic_2005}
\BIBentryALTinterwordspacing
M.~Glor, ``\BIBforeignlanguage{en}{Electrostatic ignition hazards in the process industry},'' \emph{\BIBforeignlanguage{en}{Journal of Electrostatics}}, vol.~63, no. 6-10, pp. 447--453, Jun. 2005. [Online]. Available: \url{https://linkinghub.elsevier.com/retrieve/pii/S030438860500015X}
\BIBentrySTDinterwordspacing

\bibitem{wilms_ml_2024}
\BIBentryALTinterwordspacing
C.~Wilms, W.~Xu, G.~Ozler, S.~Jantač, S.~Schmelter, and H.~Grosshans, ``\BIBforeignlanguage{en}{{ML} enhanced measurement of the electrostatic charge distribution of powder conveyed through a duct},'' \emph{\BIBforeignlanguage{en}{Journal of Loss Prevention in the Process Industries}}, vol.~92, p. 105474, Dec. 2024. [Online]. Available: \url{https://linkinghub.elsevier.com/retrieve/pii/S0950423024002328}
\BIBentrySTDinterwordspacing

\bibitem{deng_electrostatic_2023}
\BIBentryALTinterwordspacing
T.~Deng, V.~Garg, and M.~S.~A. Bradley, ``\BIBforeignlanguage{en}{Electrostatic {Charging} of {Fine} {Powders} and {Assessment} of {Charge} {Polarity} {Using} an {Inductive} {Charge} {Sensor}},'' \emph{\BIBforeignlanguage{en}{Nanomanufacturing}}, vol.~3, no.~3, pp. 281--292, Sep. 2023, number: 3 Publisher: Multidisciplinary Digital Publishing Institute. [Online]. Available: \url{https://www.mdpi.com/2673-687X/3/3/18}
\BIBentrySTDinterwordspacing

\bibitem{murtomaa_electrostatic_2003}
\BIBentryALTinterwordspacing
M.~Murtomaa, E.~Räsänen, J.~Rantanen, A.~Bailey, E.~Laine, J.-P. Mannermaa, and J.~Yliruusi, ``\BIBforeignlanguage{en}{Electrostatic measurements on a miniaturized fluidized bed},'' \emph{\BIBforeignlanguage{en}{Journal of Electrostatics}}, vol.~57, no.~1, pp. 91--106, Jan. 2003. [Online]. Available: \url{https://linkinghub.elsevier.com/retrieve/pii/S0304388602001213}
\BIBentrySTDinterwordspacing

\bibitem{heinert_decay_2022}
\BIBentryALTinterwordspacing
C.~Heinert, R.~M. Sankaran, and D.~J. Lacks, ``\BIBforeignlanguage{en}{Decay of electrostatic charge on surfaces due solely to gas phase interactions},'' \emph{\BIBforeignlanguage{en}{Journal of Electrostatics}}, vol. 115, p. 103663, Jan. 2022. [Online]. Available: \url{https://linkinghub.elsevier.com/retrieve/pii/S0304388621001078}
\BIBentrySTDinterwordspacing

\bibitem{jantac_triboelectric_2025}
\BIBentryALTinterwordspacing
S.~Jantač, J.~Pelcová, J.~Sklenářová, M.~Drápela, H.~Grosshans, and J.~Kosek, ``\BIBforeignlanguage{en}{Triboelectric charging model for particles with rough surfaces},'' \emph{\BIBforeignlanguage{en}{Advanced Powder Technology}}, vol.~36, no.~3, p. 104787, Mar. 2025. [Online]. Available: \url{https://linkinghub.elsevier.com/retrieve/pii/S0921883125000081}
\BIBentrySTDinterwordspacing

\bibitem{grunebeck_size_2024}
\BIBentryALTinterwordspacing
C.~Grünebeck, F.~Chioma Onyeagusi, J.~Teiser, and G.~Wurm, ``\BIBforeignlanguage{en}{Size dependent polarities in tribocharged dust aggregates},'' \emph{\BIBforeignlanguage{en}{Soft Matter}}, vol.~20, no.~48, pp. 9572--9577, 2024, publisher: Royal Society of Chemistry. [Online]. Available: \url{https://pubs.rsc.org/en/content/articlelanding/2024/sm/d4sm01013b}
\BIBentrySTDinterwordspacing

\bibitem{xu_experimental_2023}
\BIBentryALTinterwordspacing
W.~Xu and H.~Grosshans, ``\BIBforeignlanguage{en}{Experimental study of humidity influence on triboelectric charging of particle-laden duct flows},'' \emph{\BIBforeignlanguage{en}{Journal of Loss Prevention in the Process Industries}}, vol.~81, p. 104970, Feb. 2023. [Online]. Available: \url{https://linkinghub.elsevier.com/retrieve/pii/S0950423022002467}
\BIBentrySTDinterwordspacing

\bibitem{cruise_triboelectric_2023}
\BIBentryALTinterwordspacing
R.~D. Cruise, S.~O. Starr, K.~Hadler, and J.~J. Cilliers, ``\BIBforeignlanguage{en}{Triboelectric charge saturation on single and multiple insulating particles in air and vacuum},'' \emph{\BIBforeignlanguage{en}{Scientific Reports}}, vol.~13, no.~1, p. 15178, Sep. 2023, number: 1 Publisher: Nature Publishing Group. [Online]. Available: \url{https://www.nature.com/articles/s41598-023-42265-0}
\BIBentrySTDinterwordspacing

\bibitem{alfano_computational_2021}
\BIBentryALTinterwordspacing
F.~O. Alfano, A.~Di~Renzo, F.~P. Di~Maio, and M.~Ghadiri, ``\BIBforeignlanguage{en}{Computational analysis of triboelectrification due to aerodynamic powder dispersion},'' \emph{\BIBforeignlanguage{en}{Powder Technology}}, vol. 382, pp. 491--504, Apr. 2021. [Online]. Available: \url{https://linkinghub.elsevier.com/retrieve/pii/S0032591021000188}
\BIBentrySTDinterwordspacing

\bibitem{puttick_avoidance_2008}
S.~Puttick, ``\BIBforeignlanguage{en}{Avoidance of ignition sources as a basis of safety – limitations and challenges},'' \emph{\BIBforeignlanguage{en}{Institution of Chemical Engineers; Symposium Series}}, no. 154, 2008.

\bibitem{choi_experimental_2015}
\BIBentryALTinterwordspacing
K.~Choi, K.~Choi, and K.~Nishimura, ``\BIBforeignlanguage{en}{Experimental study on the influence of the nitrogen concentration in the air on the minimum ignition energies of combustible powders due to electrostatic discharges},'' \emph{\BIBforeignlanguage{en}{Journal of Loss Prevention in the Process Industries}}, vol.~34, pp. 163--166, Mar. 2015. [Online]. Available: \url{https://linkinghub.elsevier.com/retrieve/pii/S0950423015000467}
\BIBentrySTDinterwordspacing

\bibitem{aneke_potential_2015}
\BIBentryALTinterwordspacing
M.~Aneke and M.~Wang, ``\BIBforeignlanguage{en}{Potential for improving the energy efficiency of cryogenic air separation unit ({ASU}) using binary heat recovery cycles},'' \emph{\BIBforeignlanguage{en}{Applied Thermal Engineering}}, vol.~81, pp. 223--231, Apr. 2015. [Online]. Available: \url{https://linkinghub.elsevier.com/retrieve/pii/S1359431115001428}
\BIBentrySTDinterwordspacing

\bibitem{ebadat_testing_1996}
\BIBentryALTinterwordspacing
V.~Ebadat and J.~C. Mulligan, ``\BIBforeignlanguage{en}{Testing the suitability of {FIBCs} for use in flammable atmospheres},'' \emph{\BIBforeignlanguage{en}{Process Safety Progress}}, vol.~15, no.~3, pp. 123--127, Sep. 1996. [Online]. Available: \url{https://aiche.onlinelibrary.wiley.com/doi/10.1002/prs.680150304}
\BIBentrySTDinterwordspacing

\bibitem{ohara_faraday_2025}
\BIBentryALTinterwordspacing
T.~F. O’Hara, D.~P. Reid, G.~L. Marsden, and K.~L. Aplin, ``\BIBforeignlanguage{en}{Faraday cup measurements of triboelectrically charged granular material: a modular interpretation methodology},'' \emph{\BIBforeignlanguage{en}{Soft Matter}}, 2025. [Online]. Available: \url{https://xlink.rsc.org/?DOI=D4SM01124D}
\BIBentrySTDinterwordspacing

\bibitem{houghton_triboelectric_2013}
\BIBentryALTinterwordspacing
I.~M.~P. Houghton, K.~L. Aplin, and K.~A. Nicoll, ``\BIBforeignlanguage{en}{Triboelectric {Charging} of {Volcanic} {Ash} from the 2011 {Grímsvötn} {Eruption}},'' \emph{\BIBforeignlanguage{en}{Physical Review Letters}}, vol. 111, no.~11, p. 118501, Sep. 2013. [Online]. Available: \url{https://link.aps.org/doi/10.1103/PhysRevLett.111.118501}
\BIBentrySTDinterwordspacing

\bibitem{aplin_self-calibrating_2001}
\BIBentryALTinterwordspacing
K.~L. Aplin and R.~G. Harrison, ``\BIBforeignlanguage{en}{A self-calibrating programable mobility spectrometer for atmospheric ion measurements},'' \emph{\BIBforeignlanguage{en}{Review of Scientific Instruments}}, vol.~72, no.~8, pp. 3467--3469, Aug. 2001. [Online]. Available: \url{https://pubs.aip.org/rsi/article/72/8/3467/346700/A-self-calibrating-programable-mobility}
\BIBentrySTDinterwordspacing

\bibitem{keithley_low_1998}
J.~F. Keithley, \emph{Low {Level} {Measurements}. {Precision} {DC} {Current}, {Voltage} and {Resistance} {Measurements}}.\hskip 1em plus 0.5em minus 0.4em\relax Keithley Instruments, 1998.

\bibitem{sun_functionalized_2021}
\BIBentryALTinterwordspacing
J.~Sun, K.~Tu, S.~Büchele, S.~M. Koch, Y.~Ding, S.~N. Ramakrishna, S.~Stucki, H.~Guo, C.~Wu, T.~Keplinger, J.~Pérez-Ramírez, I.~Burgert, and G.~Panzarasa, ``\BIBforeignlanguage{English}{Functionalized wood with tunable tribopolarity for efficient triboelectric nanogenerators},'' \emph{\BIBforeignlanguage{English}{Matter}}, vol.~4, no.~9, pp. 3049--3066, Sep. 2021, publisher: Elsevier. [Online]. Available: \url{https://www.cell.com/matter/abstract/S2590-2385(21)00393-3}
\BIBentrySTDinterwordspacing

\bibitem{antony_untreated_2025}
\BIBentryALTinterwordspacing
L.~Antony, A.~Giuri, R.~Mastria, E.~Kovalska, J.~Kirkwood, A.~Danaa, S.~Russo, M.~F. Craciun, and A.~Rizzo, ``\BIBforeignlanguage{en}{Untreated {Natural} {Wood}-{Based} {Triboelectric} {Nanogenerator} for {Floor} {Charge} {Energy} {Harvesting}},'' \emph{\BIBforeignlanguage{en}{Advanced Sustainable Systems}}, vol.~9, no.~1, p. 2400493, 2025, \_eprint: https://advanced.onlinelibrary.wiley.com/doi/pdf/10.1002/adsu.202400493. [Online]. Available: \url{https://onlinelibrary.wiley.com/doi/abs/10.1002/adsu.202400493}
\BIBentrySTDinterwordspacing

\bibitem{ohara_surface_2025}
\BIBentryALTinterwordspacing
T.~F. O'Hara, E.~Player, G.~Ackroyd, P.~J. Caine, and K.~L. Aplin, ``Surface {Material} {Dependence} in {Powder} {Triboelectric} {Charging},'' 2025. [Online]. Available: \url{https://materialsdatafacility.org/detail/566d3ffd-b098-408a-93f7-fc4a826433b8-1.0}
\BIBentrySTDinterwordspacing

\bibitem{tom_ohara_vb22224surface-material-dependence--powder-triboelectric-charging_2025}
\BIBentryALTinterwordspacing
T.~O'Hara, ``vb22224/{Surface}-{Material}-{Dependence}-in-{Powder}-{Triboelectric}-{Charging}: {Surface} {Material} {Dependence} in {Powder} {Triboelectric} {Charging},'' Jul. 2025. [Online]. Available: \url{https://zenodo.org/doi/10.5281/zenodo.15825255}
\BIBentrySTDinterwordspacing

\end{thebibliography}

\end{document}